\documentclass{moriond}

\def\to{\rightarrow}

\newcommand{\Photo}{}

\begin{document}
\vspace*{4cm}
\title{pMSSM SUSY and the ATLAS Z+jets+MET Excess}

\author{ Thomas G. Rizzo }

\address{SLAC National Accelerator Laboratory,\\
Menlo Park, CA, USA}

\maketitle\abstracts
{A possible explanation of the excesses observed by ATLAS in the Z+jets+MET 
channel within the p(henomenological)MSSM is discussed. We have found that 
the cascade of first/second generation squarks to binos to Higgsino LSPs, with 
judiciously chosen sparticle masses, can provide a viable scenario that is 
consistent with all other experimental constraints.}

The Standard Model (SM) is obviously incomplete and new physics is expected to show up at some point 
to address the many questions that it leaves unanswered. Supersymmetry (SUSY) provides one of the 
leading candidate frameworks for physics beyond the SM and new SUSY searches are continuing at the 
13 TeV, Run 2 LHC and elsewhere. At the 8 TeV LHC , ATLAS \cite{exp} observed a $3\sigma$ excess in the 
Z+(2 or more)jets+MET SUSY search channel corresponding to roughly $\sim 15-20$ additional events above 
the SM background estimate.  On the otherhand, CMS, with a somewhat different analysis which had no 
$H_T$ cut, observed an event rate consistent with SM expectations \cite{exp}. Given the differing 
kinematic requirements of the two experiments we can ask (at least) three questions: Are there MSSM 
SUSY models that are consistent with both results? What are the properties of these models? What do 
such models predict for LHC Run 2 at 13 TeV? In order to make progress addressing these questions 
we \cite{Cahill-Rowley:2015cha} began by employing some of the pMSSM model points from our own 
earlier detailed study of the associated 19-dimensional parameter space \cite{Cahill-Rowley:2014twa} 
as well as a comparable set of pMSSM model points provided by ATLAS \cite{Aad:2015baa}. Using these, 
we identified specific model points which gave a significant signal rate in the ATLAS analysis 
while simultaneously satisfying the bounds that were imposed by the null CMS search These successful 
points were then further employed as templates/seeds to examine the surrounding parameter space. 
The advantage of using these initial seed points in such a study is that they already satisfy the other 
existing LHC SUSY searches as well as the constraints from flavor physics, precision measurements and 
dark matter(DM) search experiments.

\begin{figure}[htpb]
\centerline{\includegraphics[width=2.9in,angle=0]{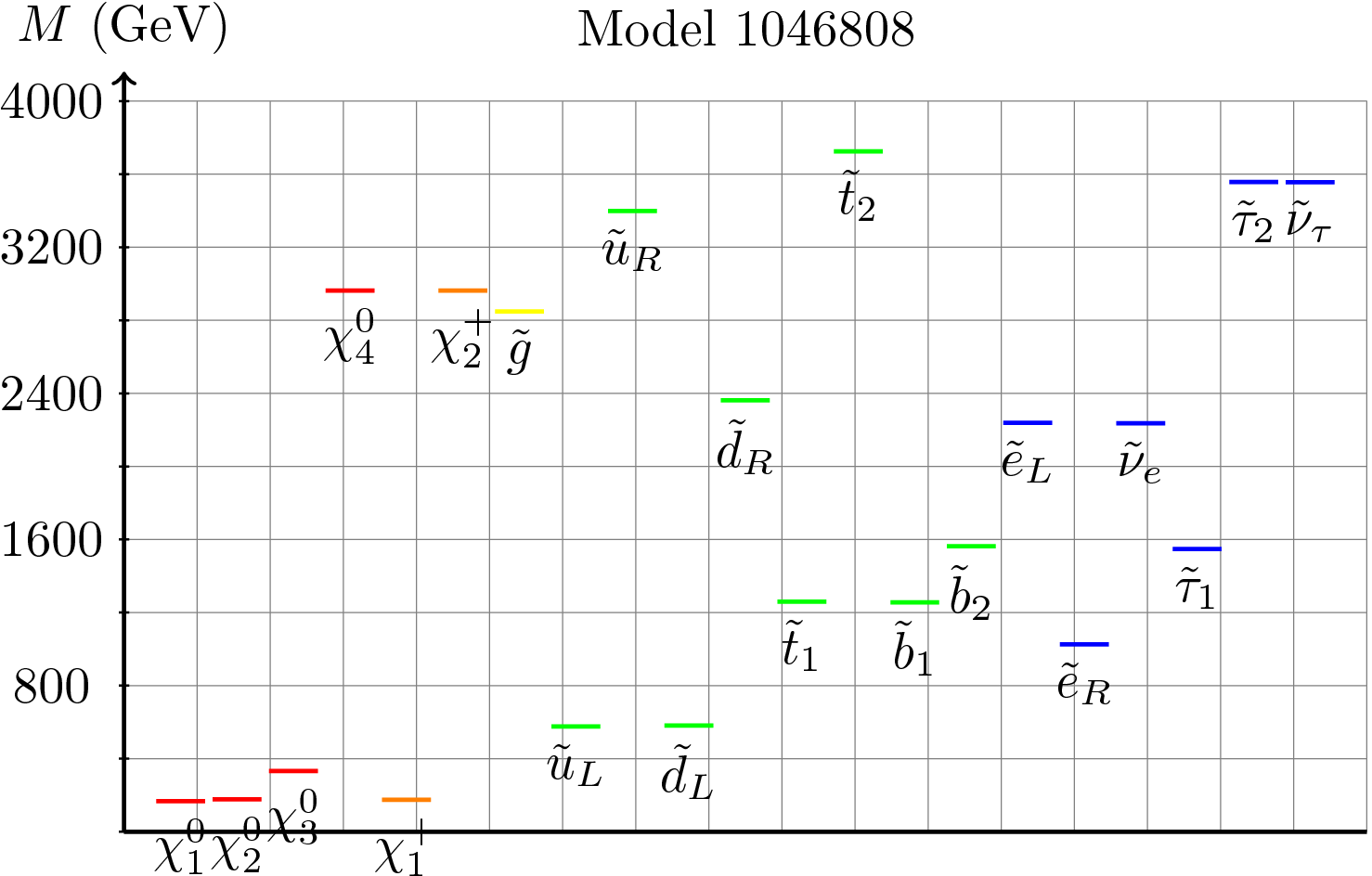}
\hspace {0.7cm}
\includegraphics[width=2.6in,angle=0]{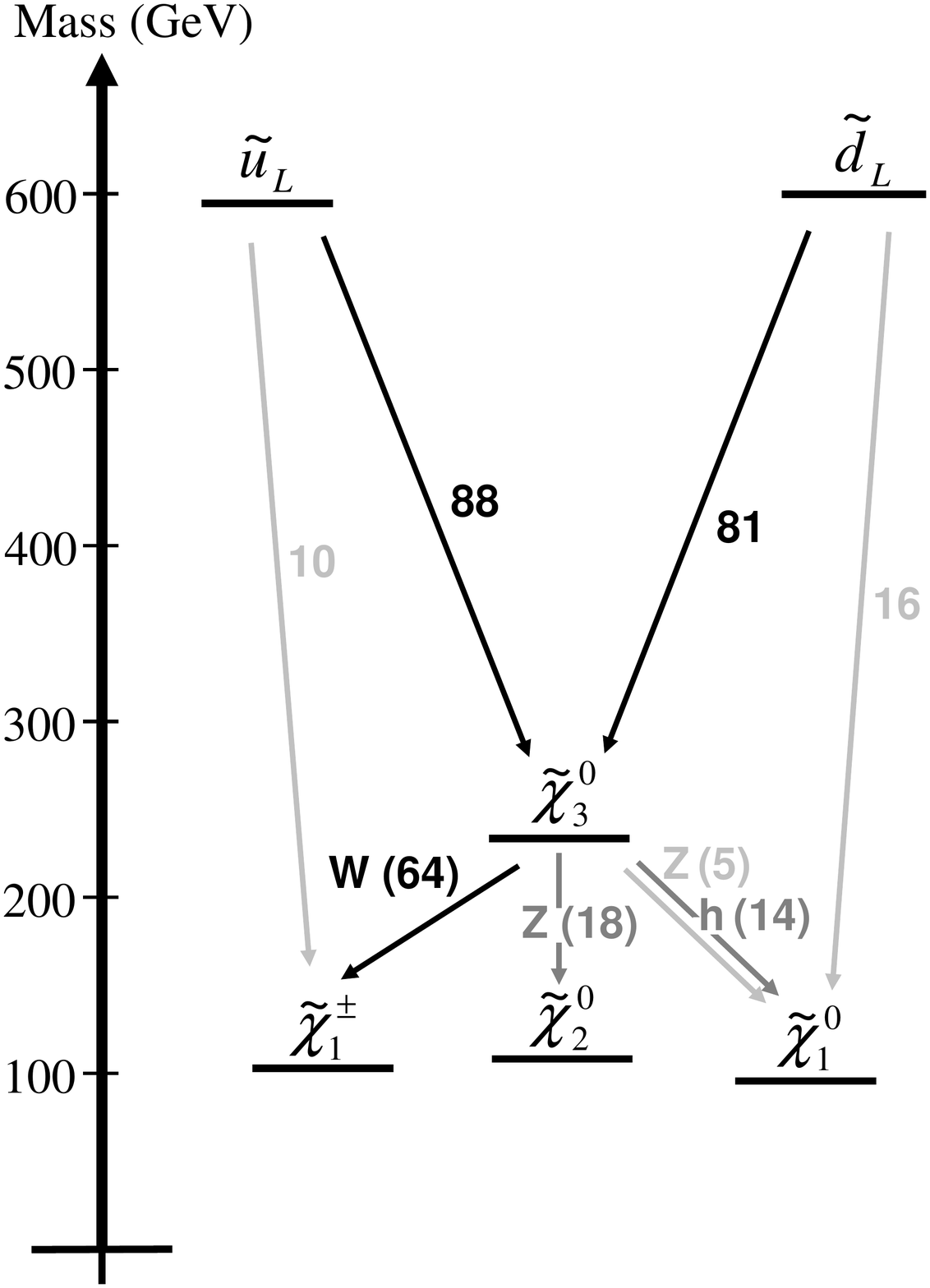}}
\vspace*{0.50cm}
\caption{(Left) Sample seed model spectrum and (Right) the corresponding squark to bino to Higgsino 
decay pattern.}
\label{fig1}
\end{figure}

Before beginning this search for successful seed models, the expectation was that a SUSY cascade 
is taking place: some initial light colored sparticle (hence a QCD cross section on order to obtain 
the necessary event rate) is produced which subsequently decays to the LSP via a 2-step process 
during which the jets+Z (and possible other particles) are produced. The initial colored state must 
also be inhibited from decaying directly to the LSP so as to avoid the powerful jets+MET search 
constraints. We would also anticipate that much of the remainder of the SUSY spectrum is essentially 
decoupled from this cascade. These considerations tell us that the LSP is most likely not a bino as 
squarks of any flavor will decay to them directly. They also tell us that that any initial squarks 
are not likely from the third generation as they will decay to both binos and Higgsinos (as well as 
winos if the squarks are L-R mixed states). The LSP and intermediate color singlet state cannot be 
combinations of binos and winos as they do not couple to the Z. Further considerations along these 
lines rapidly narrow down the set of possible choices to only a few candidates. 

Indeed what we find from scanning over the set of pMSSM models as potential seed points lives up 
to these expectations.  Fig.~\ref{fig1} show a typical 
`successful' sparticle spectrum for a seed model. Here, a light $1^{st}/2^{nd}$ generation 
squark decays into an intermediate bino state by emitting a jet and then the bino itself decays 
to the lighter Higgsinos via Higgs or gauge boson emission. The bino and Higgsino are both 
slightly mixed thus allowing for these decay modes. The small bino content of the LSP suppresses 
the direct squark to LSP transition process as needed. Much of the rest of the spectrum, especially 
the gluino, is seen to be decoupled. However, even though the wino is fairly heavy here, the 
Higgsinos obtain a small wino content allowing for the squark decay to them via W emission.  
In addition to the squark doublet being the initiator of the cascade, seed points where only 
one of $\tilde u_R,\tilde d_R$ are light and filled this role were also found but they generally 
yield smaller signal event rates overall since only a single squark state is present. Note that 
since the bino can decay to the Higgs plus the LSP with a reasonable branching fraction, the 
observation of this final state is a prediction for all models in this class.     

\begin{figure}[htbp]
\centering
\includegraphics[width=0.3\textwidth]{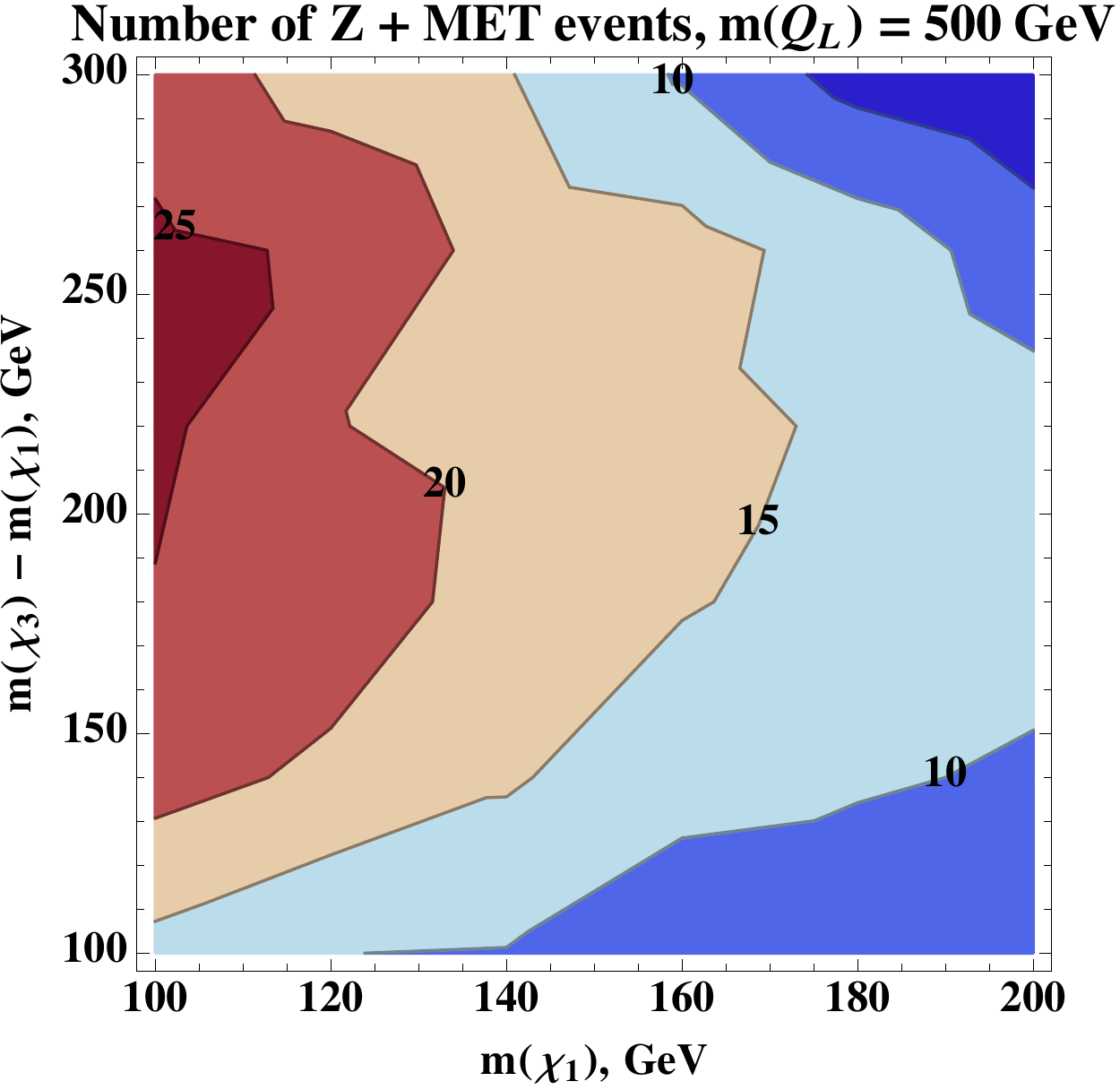}
\includegraphics[width=0.3\textwidth]{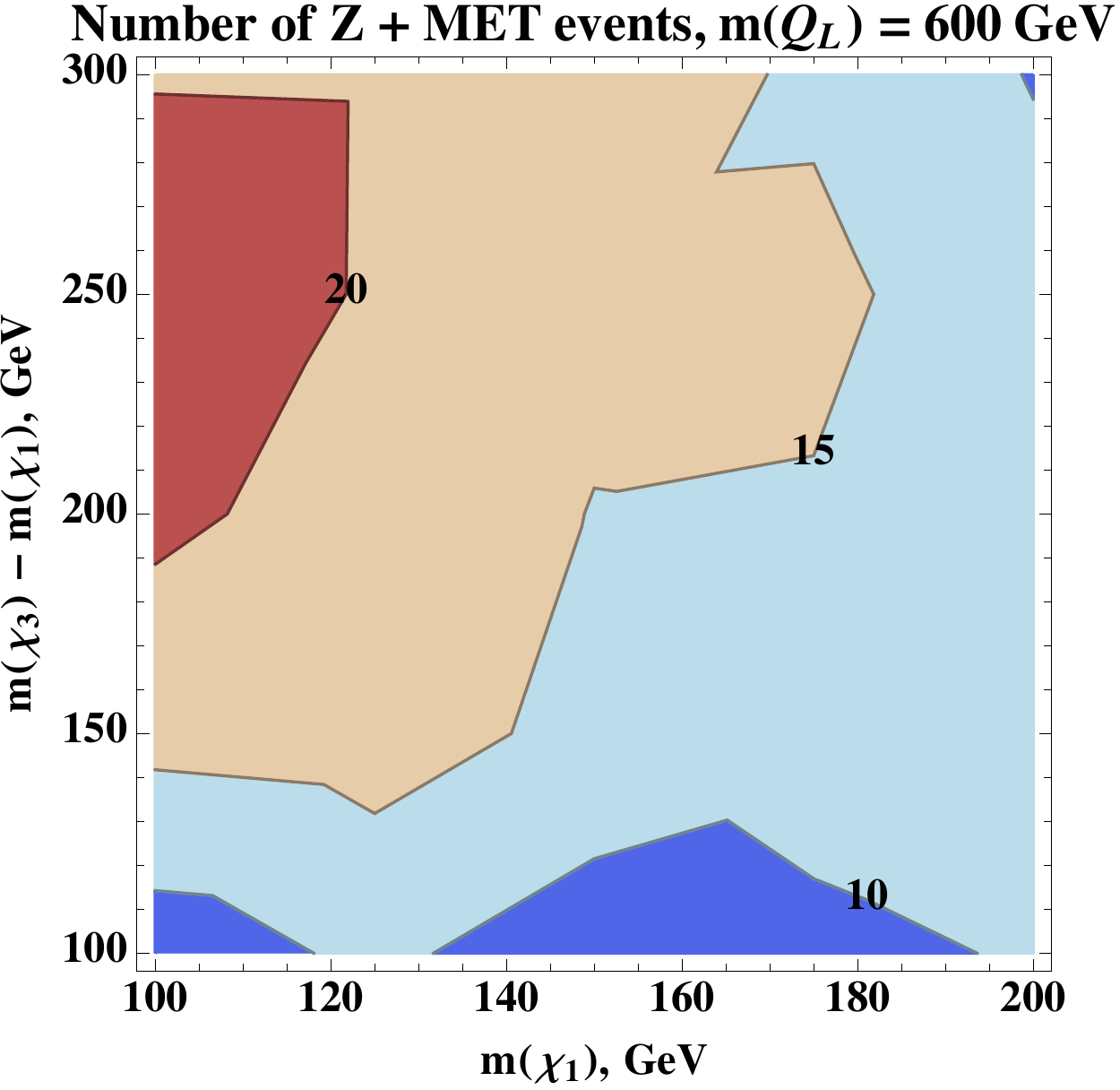}
\includegraphics[width=0.3\textwidth]{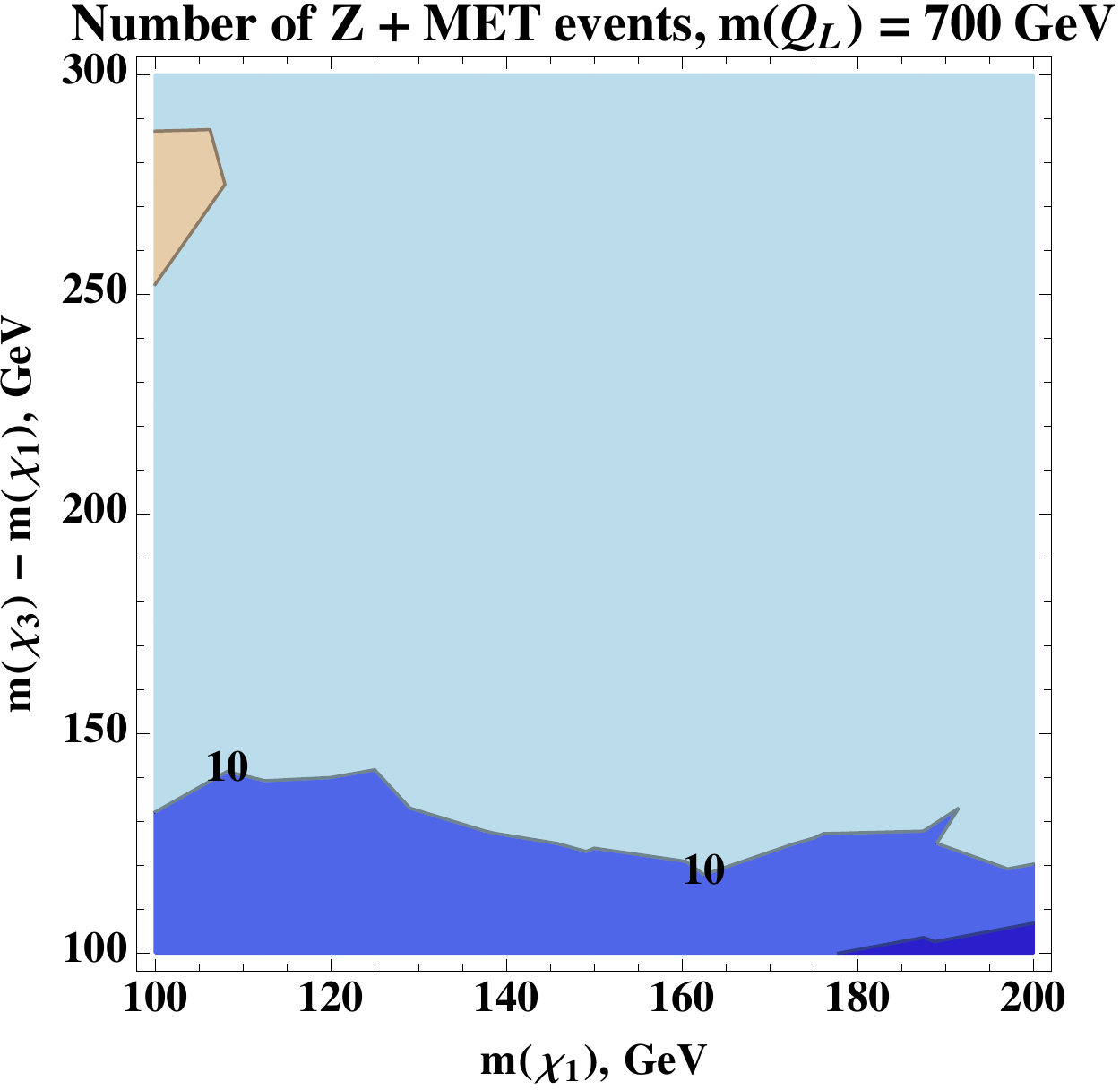}
\includegraphics[width=0.3\textwidth]{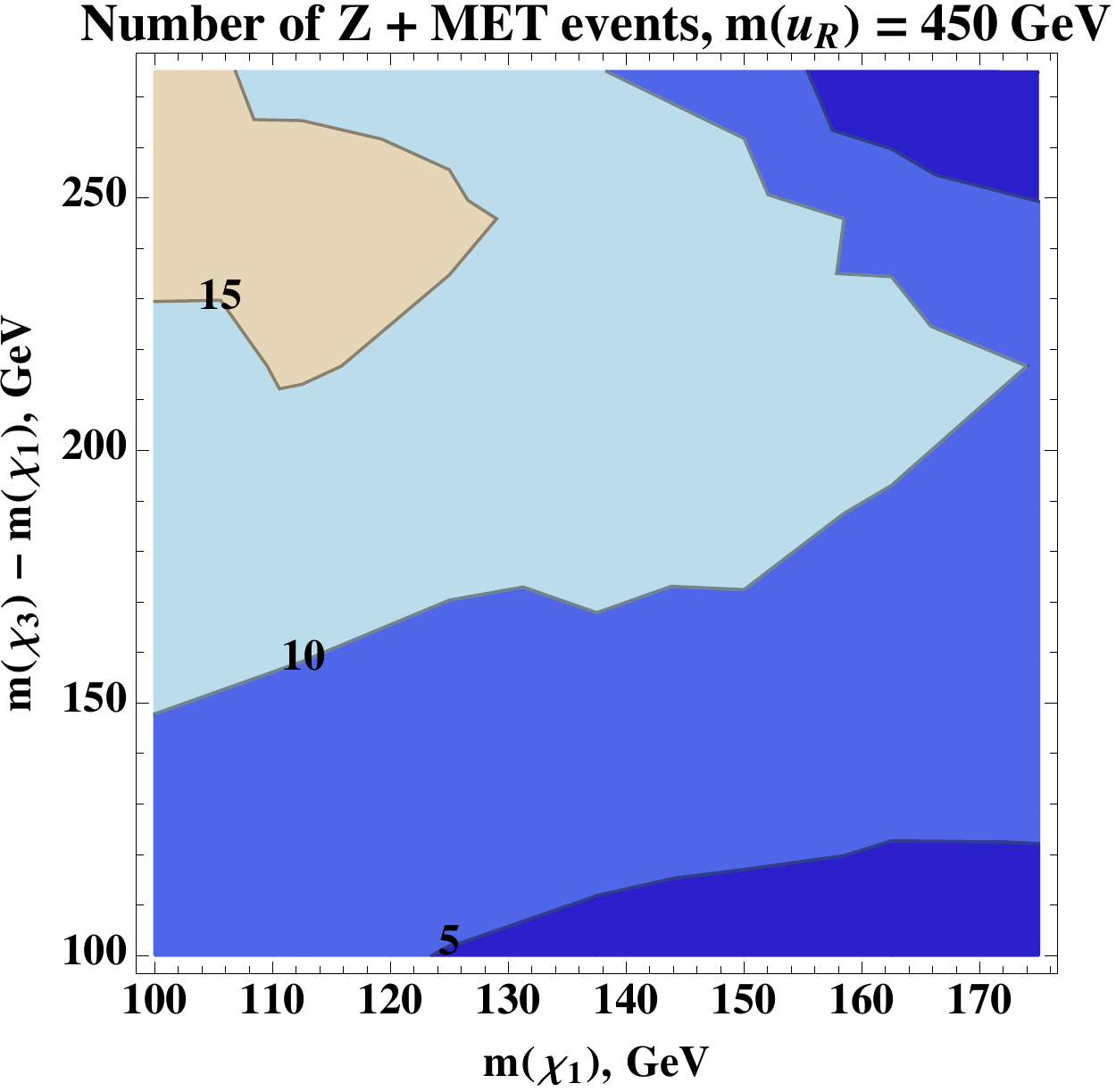}
\includegraphics[width=0.3\textwidth]{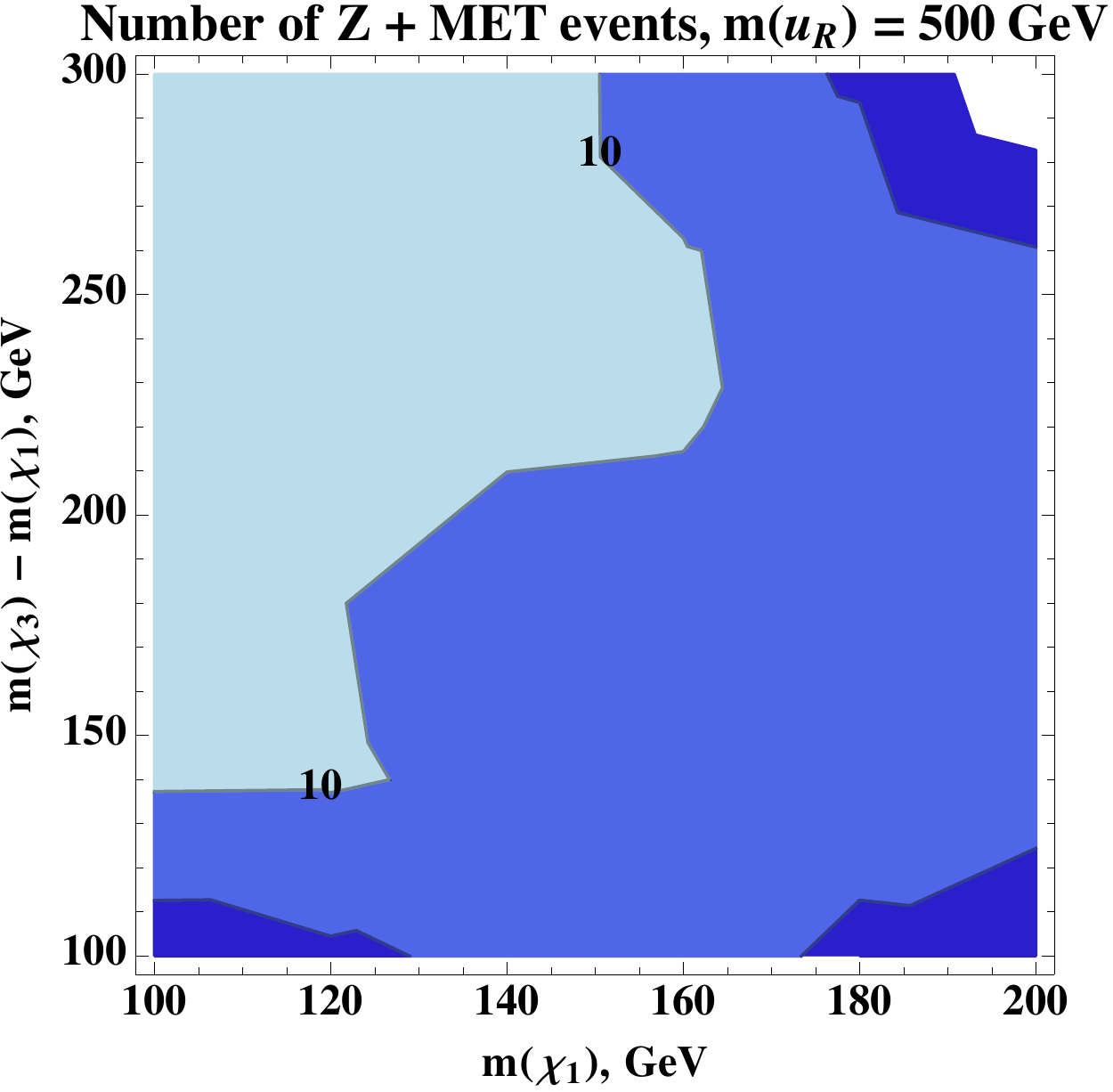}
\includegraphics[width=0.3\textwidth]{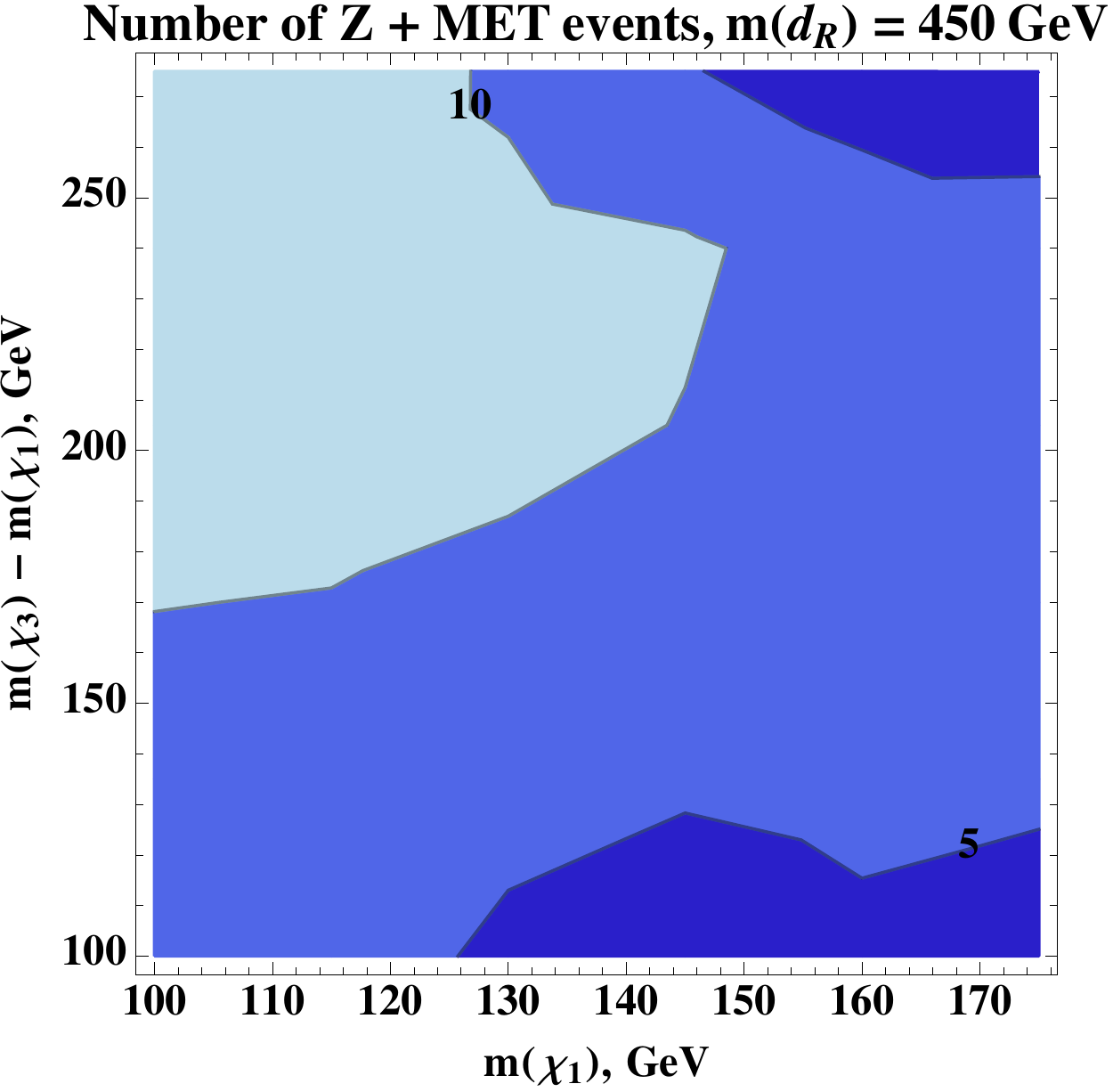}
\caption{Signal event rate contours for the ATLAS Z+jets+MET analysis in the bino-Higgsino($\chi_3^0-\chi_1^0$) 
mass difference and LSP($\chi_1^0$) mass plane. The top three panels correspond to the case of 
$\tilde Q_L=500\,, 600\,, 700$ GeV from left to right, while the bottom panels are for$\tilde u_R=450\,, 500$ GeV and 
$\tilde d_R=450$ GeV, left to right.}
\label{fig2}
\end{figure}

Given these seed points we can then move the squark, bino and Higgsino masses around in an independent  
manner and explore the impact on the Z+jets+MET rate while still requiring all the other constraints to be 
satisfied. This is essentially a simplified model analysis as it only relies on a few mass parameters 
and, in principle, the value of $\tan \beta$. The results of this scan are shown in Fig.~\ref{fig2}. 
Here we see that: ($i$) $\tilde Q_L$ initial states produce larger signal rates than do either 
$\tilde u_R$ or $\tilde d_R$, which is not unexpected. ($ii$) The preferred bino-Higgsino mass 
splitting lies above $\sim 150$ GeV; this is a bit surprising as this large mass gap allows 
for other bino decays into the Higgs final state as noted above.  ($iii$) The preferred LSP mass lies 
below $\sim 150$ GeV. Such light Higgsinos can only account for a small fraction of the thermal DM relic 
density and may eventually have a chance of being observed `directly' through the monojet/soft track 
analyses.

\begin{figure}[htbp]
\centering
\includegraphics[width=0.45\textwidth]{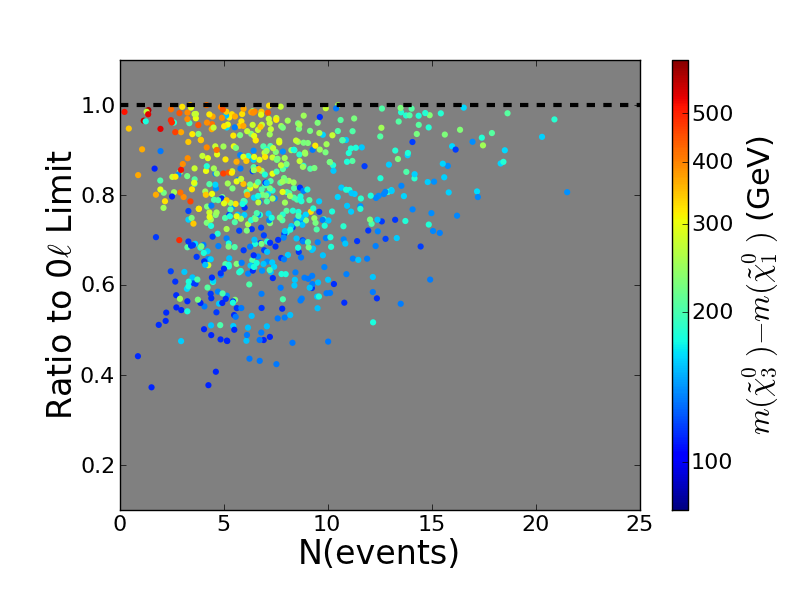}
\includegraphics[width=0.45\textwidth]{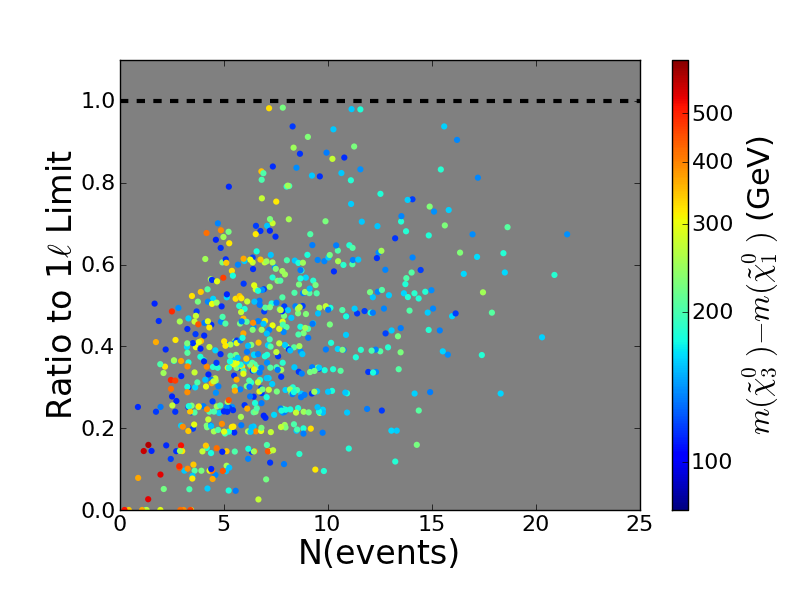}
\includegraphics[width=0.45\textwidth]{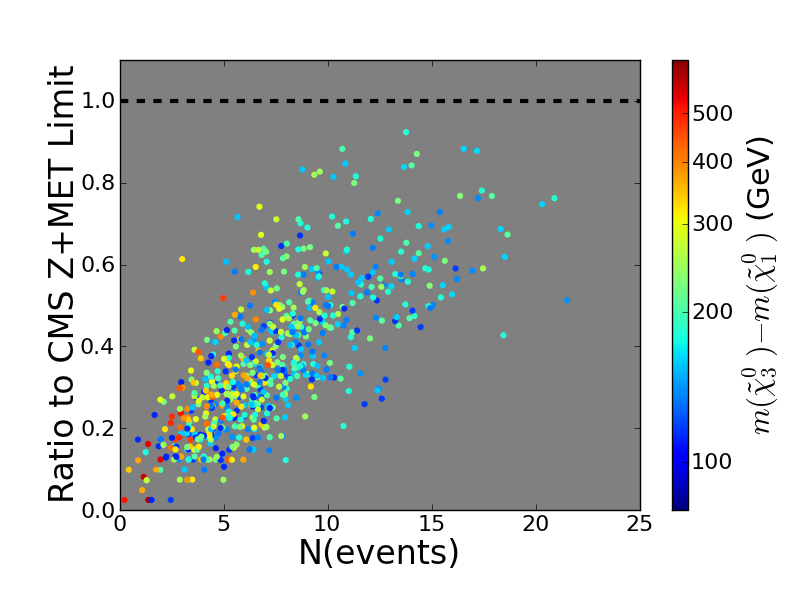}
\caption{Ratio of the predicted number of events for the $\tilde Q_L$ models in our simplified grid scan to the 
ATLAS 95\% CL. search limit for the 0l+jets and 1l+jets channels as well as that for the CMS dileption search, 
respectively, as functions of the number of predicted signal events for the ATLAS Z+jets+MET search.  The color 
code corresponds to the value of the $\chi_3^0-\chi_1^0$ mass splitting.}
\label{fig3}
\end{figure}

Moving forward we can next ask how the various ATLAS and CMS 8 TeV search requirements are shaping the 
predicted values for the number of expected Z+jets+MET events satisfying the ATLAS cuts and how `far' the 
successful models are from the constraint limits. Fig.~\ref{fig3} show these results for the case of 
the $\tilde Q_L$ initiating the cascade. The corresponding constraints for $\tilde u_R$ and $\tilde d_R$ 
are found to be qualitatively similar to these but are quantitatively weaker since, overall, these cases 
generally lead to fewer predicted signal events. From the clustering of model points near the upper 
boundary, we can observe that the 0l+jets ATLAS search is indeed constraining the maximum size of the 
Z+jets+MET excess from above. Due to the presence of leptonically decaying W's resulting from the decays 
of both the initial squarks (at least in the case of $\tilde Q_L$) as well as from the intermediate bino,
we see that the ATLAS 1l+jets search also plays some role in constraining the parameter space. However, 
this constraint is seen to be most influential for the Z+jets+MET rates of intermediate values. The last 
panel shows that the influence of the bound arising from the CMS null result on the dilepton rate is 
rather weak and that this constraint is relatively easy to satisfy once these other requirements are 
accounted for. Thus we see that we have demonstrated that the ATLAS 8 TeV excess and the corresponding 
null result from CMS at 8 TeV can be simultaneously accommodated within the pMSSM with all other search 
constraints also satisfied. These requirements point to a very particular SUSY scenario. 

\begin{figure}[htbp]
\centering
\includegraphics[width=0.45\textwidth]{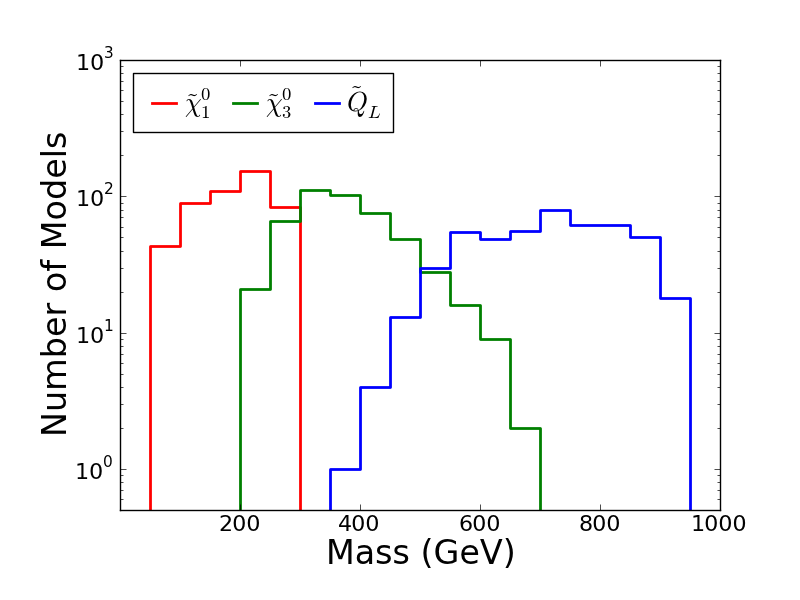}
\includegraphics[width=0.45\textwidth]{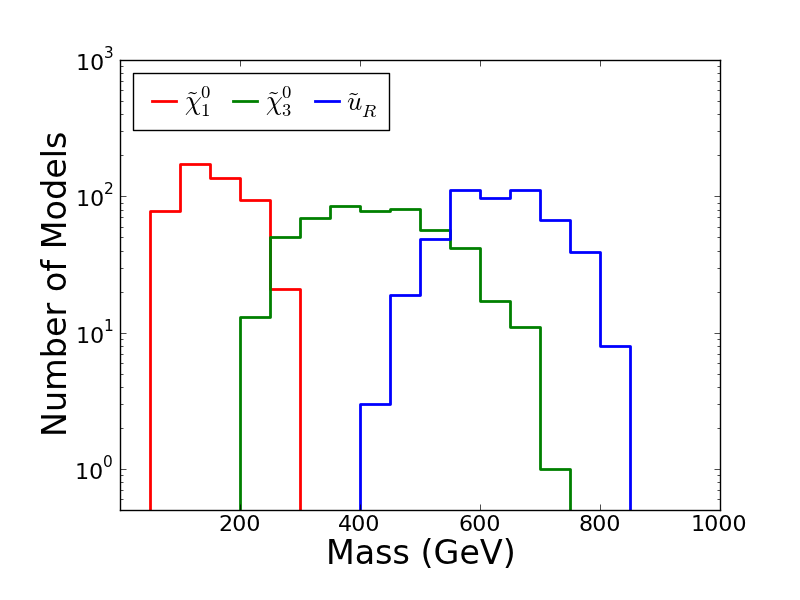}
\includegraphics[width=0.45\textwidth]{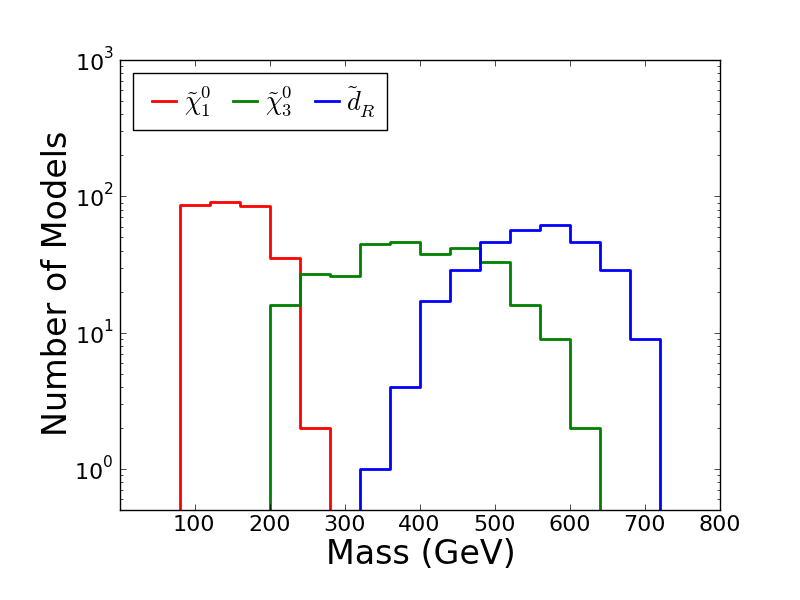}
\caption{Mass distributions of the parent squark, $\chi_3^0$ and $\chi_1^0$ states for the models from our 
grid scan that agree with all null search results and yield at least 5 events in the ATLAS $Z+$MET channel.  
The top-left, top-right, and bottom panels correspond to the parent squark cases of $\tilde Q_L$, $\tilde u_R$, 
and $\tilde d_R$, respectively.}
\label{histos}
\end{figure}

It is interesting to examine the mass distributions of the squarks, binos and Higgsinos in the subset 
of models that yield a significant Z+jets+MET signal rate; this is shown in Fig.~\ref{histos} for each 
choice of the parent squark. Overall, we see that the initial squark continues to get lighter as we go 
from $\tilde Q_L \to \tilde u_R \to \tilde d_R$ reflecting a compensation for the decrease in the production 
cross section at a fixed mass. In all three cases the Bino mass is broadly centered around $\sim 350$ GeV 
while the LSP tends to be slightly less massive in the case of the RH-squarks. This results in softer 
jets in the RH-squark case as well as slightly harder leptons arising from the Z.

\begin{figure}[htbp]
\centering
\includegraphics[width=0.3\textwidth]{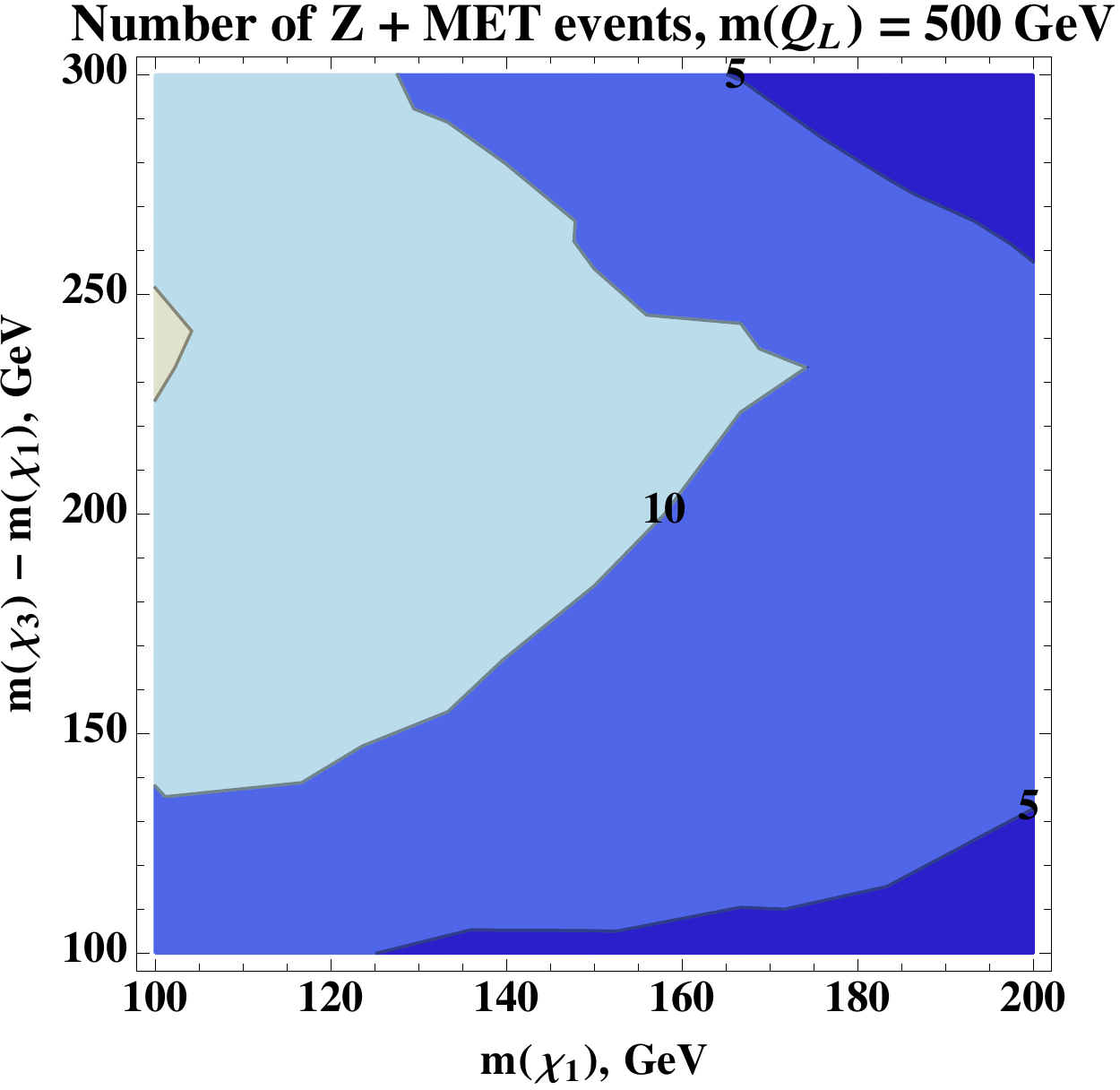}
\includegraphics[width=0.3\textwidth]{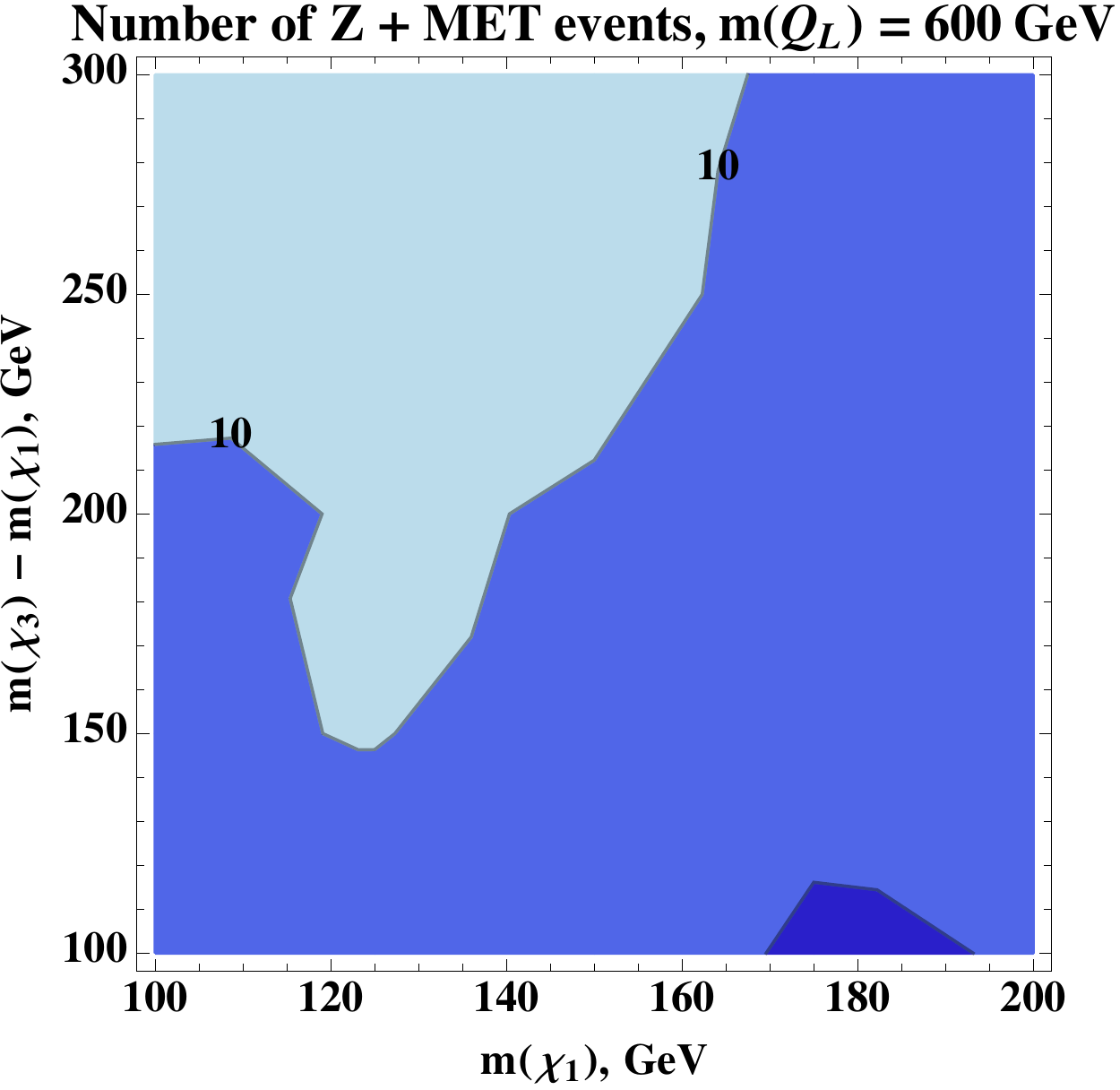}
\includegraphics[width=0.3\textwidth]{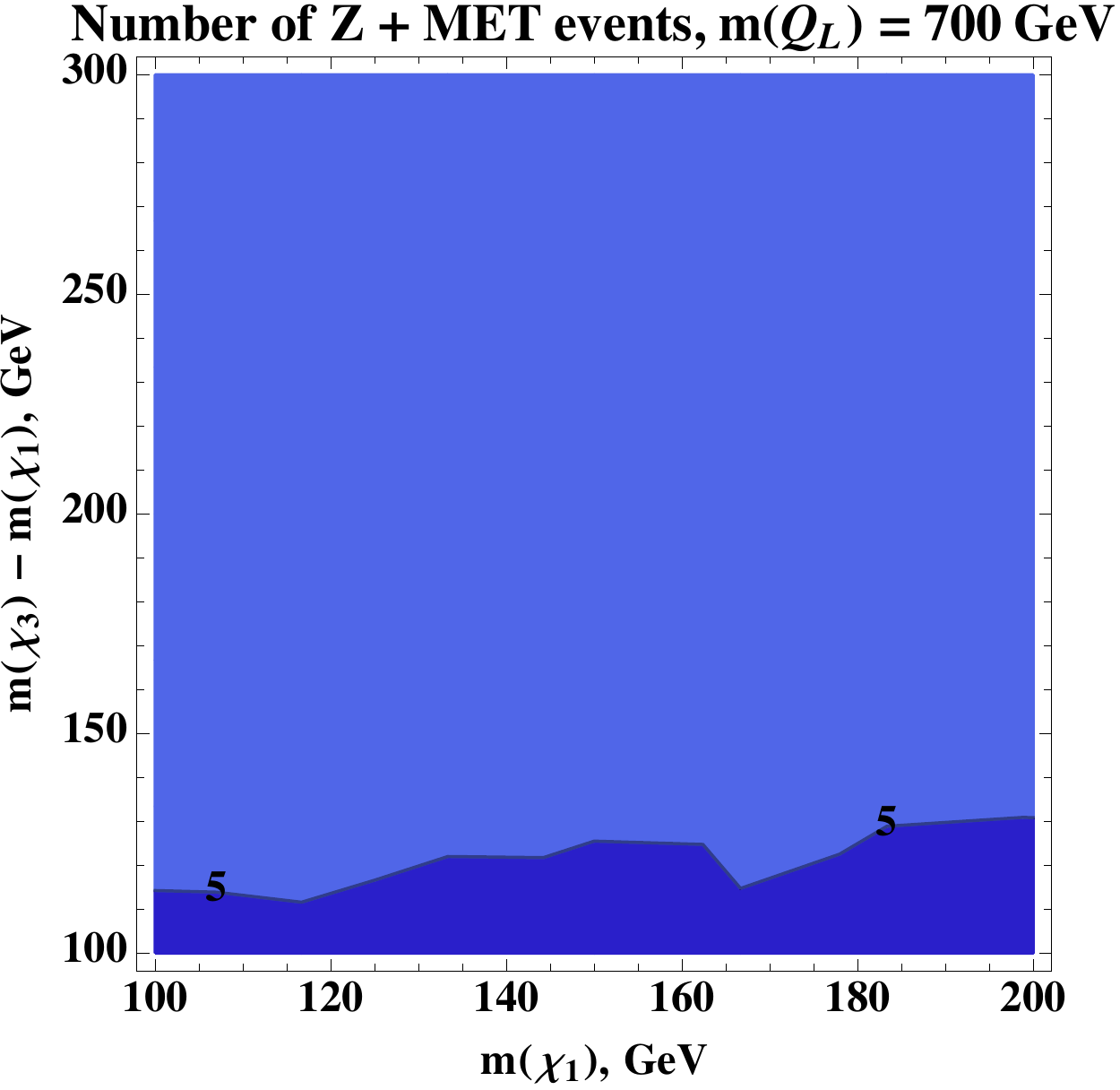}
\includegraphics[width=0.3\textwidth]{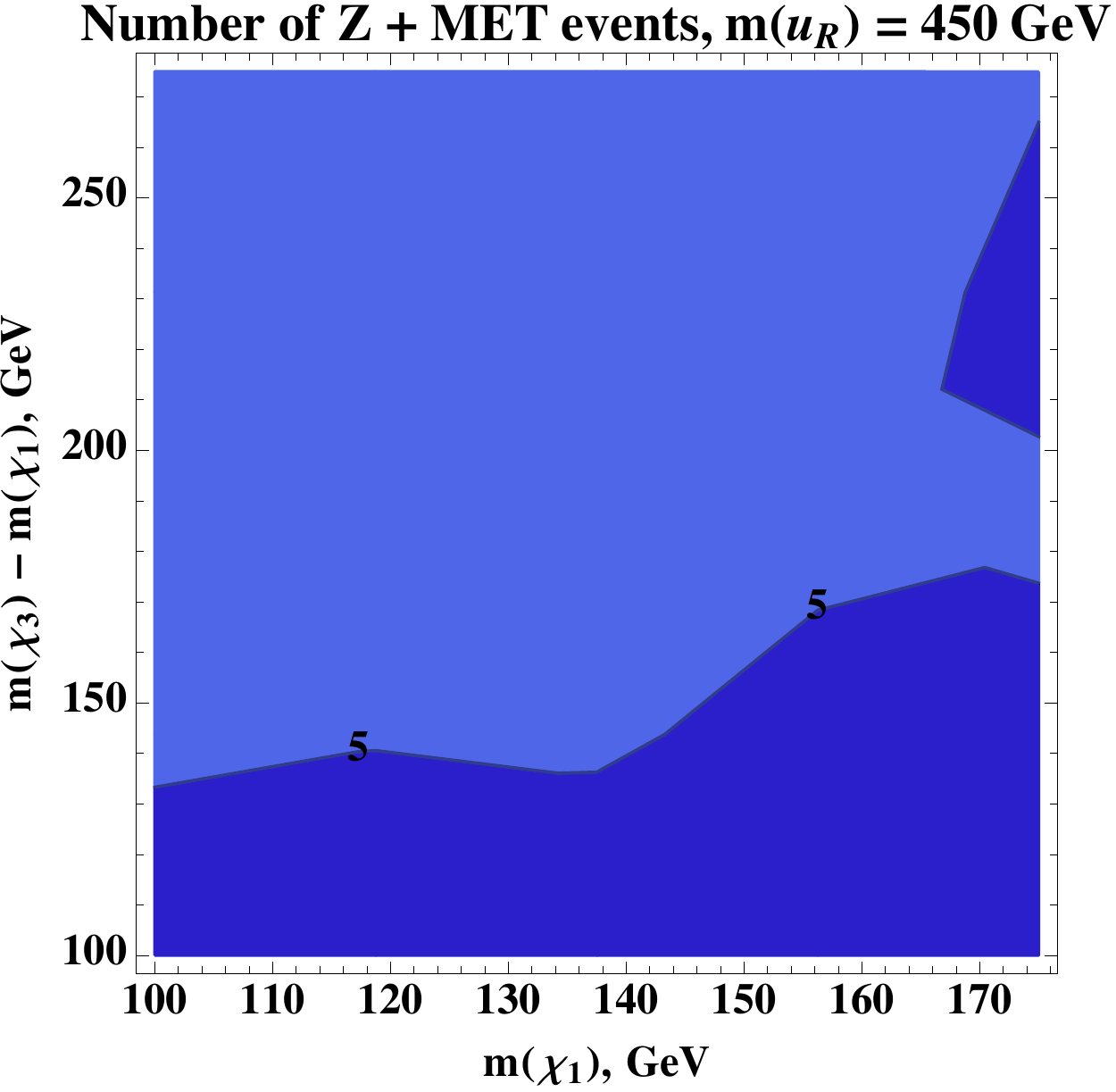}
\includegraphics[width=0.3\textwidth]{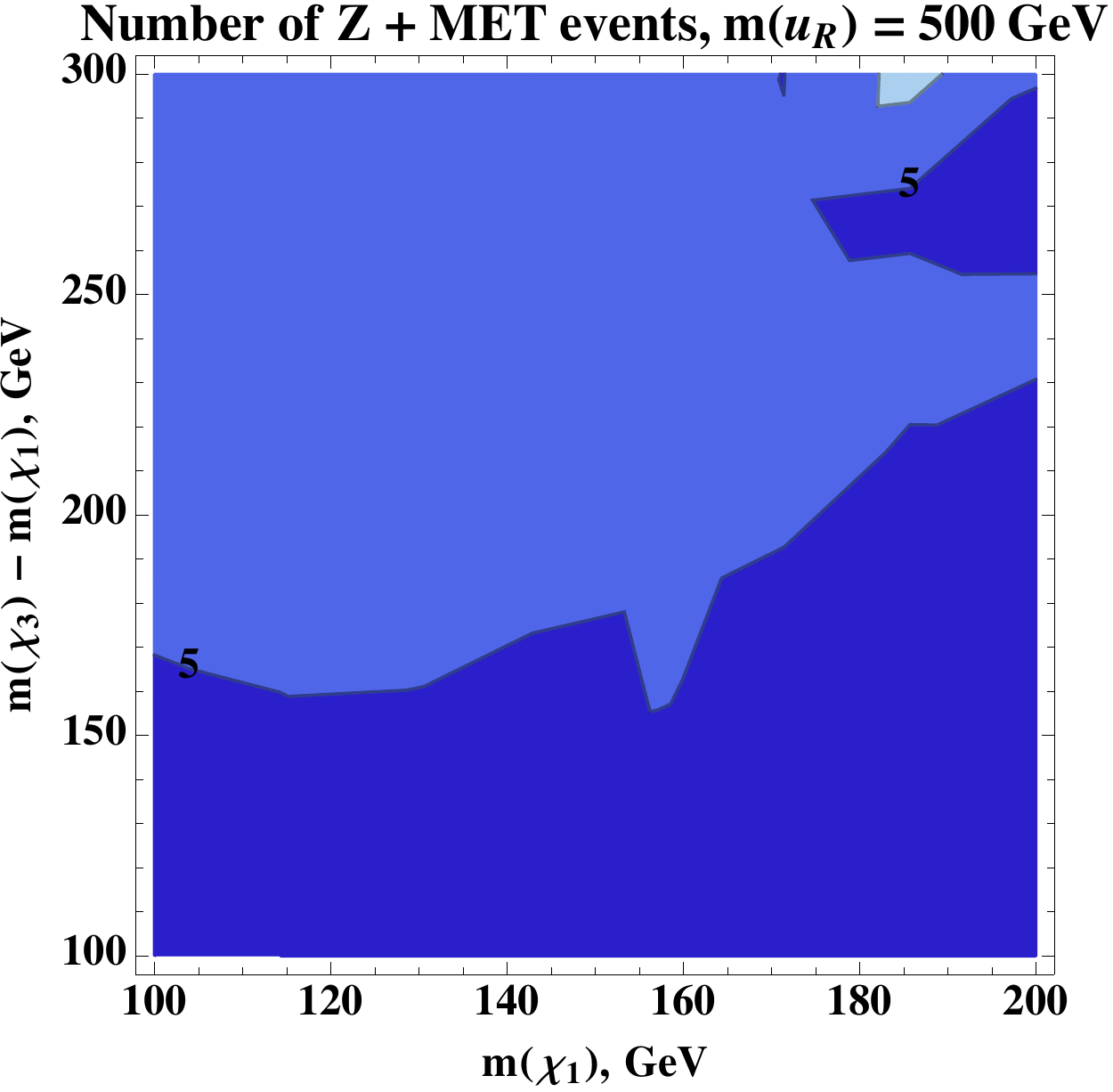}
\includegraphics[width=0.3\textwidth]{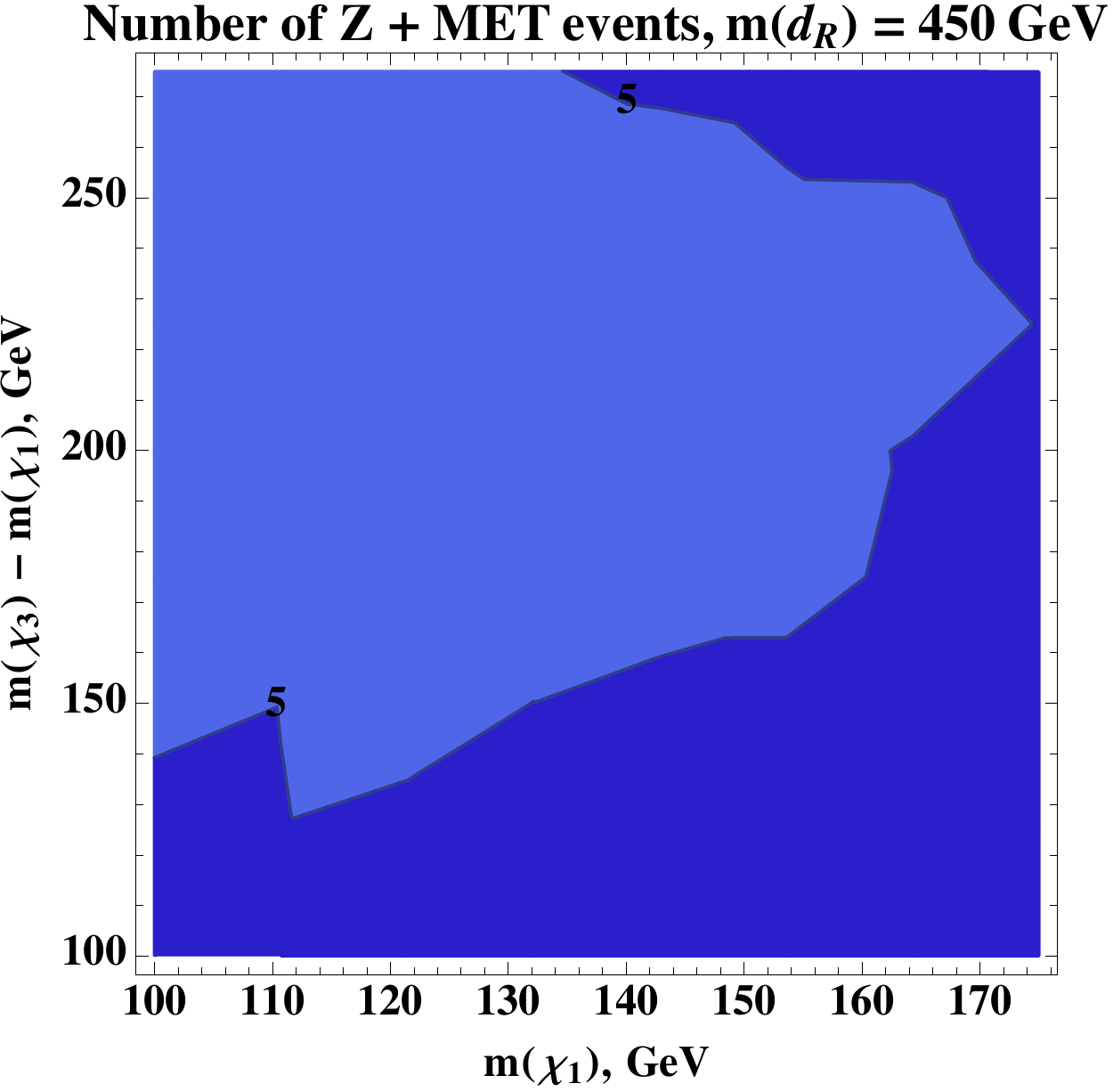}
\caption{Same as Fig.2 but now for the ATLAS 13 TeV analysis. For than 10(5) signal 
events are to be expected in the lightest(darker) region while points in the darkest regions 
lead to fewer than 5 events.}
\label{fig4}
\end{figure}

The final step in this analysis is to examine the corresponding predictions for 13 TeV. Here the 
experimental situation is a bit more confusing as ATLAS again observes a ($2\sigma$) excess of 
$\sim 10$ signal events while CMS, essentially employing the same cuts as ATLAS, still observes 
agreement with the SM expectation {\cite {exp}}. Clearly, we will need to wait until more data 
is available to clarify this situation. Fig.~\ref{fig4} shows the predicted number of 
events passing the ATLAS 13 TeV analysis in the same plane as that employed above for the case 
of the 8 TeV analysis. Due to the larger $\sqrt s$ but much smaller luminosity, fewer numbers 
of signal events are generally to be expected to be produced here. However, we do see from this 
Figure that it is possible to obtain $\sim 10$ signal events arising from $\tilde Q_L$ production 
and decay with masses of order $\sim 500-600$ GeV or more. Unfortunately the predicted rates 
arising from either $\tilde u_R$ or $\tilde d_R$ production cannot make it up to this signal 
rate threshold. As a final step we must investigate how the other ATLAS SUSY searches at 13 TeV 
are impacting on the parameter space that allows for a significant Z+jets+MET excess as this 
energy. This is work that is now in progress {\cite {new}}.

\section*{Acknowledgments}

This work was supported by the Department of Energy, Contract DE-AC02-76SF00515.


\section*{References}

\begin{thebibliography}{99}

\bibitem{exp}
  G.~Aad {\it et al.} [ATLAS Collaboration],
  Eur.\ Phys.\ J.\ C {\bf 75}, no. 7, 318 (2015)
  Erratum: [Eur.\ Phys.\ J.\ C {\bf 75}, no. 10, 463 (2015)]
  doi:10.1140/epjc/s10052-015-3661-9, 10.1140/epjc/s10052-015-3518-2
  [arXiv:1503.03290 [hep-ex]];
  V.~Khachatryan {\it et al.} [CMS Collaboration],
  JHEP {\bf 1504}, 124 (2015)
  doi:10.1007/JHEP04(2015)124
  [arXiv:1502.06031 [hep-ex]];
ATLAS Collaboration analysis note, ATLAS-CONF-2015-082 (2015);
CMS Collaboration analysis note CMS-PAS-SUS-15-011 (2015).  


\bibitem{Cahill-Rowley:2015cha} 
  M.~Cahill-Rowley, J.~L.~Hewett, A.~Ismail and T.~G.~Rizzo,
  Phys.\ Rev.\ D {\bf 92}, 075029 (2015)
  doi:10.1103/PhysRevD.92.075029
  [arXiv:1506.05799 [hep-ph]].
  
\bibitem{Cahill-Rowley:2014twa} 
  M.~Cahill-Rowley, J.~L.~Hewett, A.~Ismail and T.~G.~Rizzo,
  Phys.\ Rev.\ D {\bf 91}, no. 5, 055002 (2015)
  doi:10.1103/PhysRevD.91.055002
  [arXiv:1407.4130 [hep-ph]].

\bibitem{Aad:2015baa} 
  G.~Aad {\it et al.} [ATLAS Collaboration],
  JHEP {\bf 1510}, 134 (2015)
  doi:10.1007/JHEP10(2015)134
  [arXiv:1508.06608 [hep-ex]].

\bibitem{new}

J.L. Hewett, A. Ismail, T.G. Rizzo and T.D. Rueter, to appear. 

\end{thebibliography}
\end{document}